\documentclass{elsarticle}

\usepackage{lineno,hyperref}
\modulolinenumbers[5]

\journal{Journal of \LaTeX\ Templates}

\usepackage{xcolor}
\usepackage{amsmath}  
\usepackage{amsfonts} 
\usepackage{graphicx} 
\usepackage[utf8]{inputenc}

\begin{document}

\begin{frontmatter}

\title{Propagation of radical ideas in societies with memory:
competition between radical strength and social cohesion} 
  
\author{Andr\'es Vallejo}
\author{Arturo C. Marti}
\address{Instituto de Física, Universidad 
de la Rep\'ublica, Montevideo, Uruguay}
\date{\today}
\begin{abstract} 
Based on a dynamical model with fractional derivatives we show that
the evolution of radical groups in a society with a memory depends
mainly on the active proselytism of radical groups and the level of
cohesion in social networks. We find the conditions that determine
that the radical group either disappears or, on the contrary, achieves
a permanent significant expression.  We also discuss the possible
intervention strategies on the susceptible population and show the
conditions that lead to the disappearance of radical groups. We see
that the higher the level of memory, the maximum proportion of
radicals decreases, but those groups manage to maintain their presence
in society during a larger period of time.
\end{abstract}

\end{frontmatter}

\bibliographystyle{elsarticle-num}

\section{\label{sec:level1}Introduction}

Since the appearance of the well-known SIR
(susceptible-infected-recovered/ removed) model for the spread of
epidemics proposed by Kermack and McKendrick in 1927 \cite{Kermack},
compartmental models have become so popular that they are now, as a
first approximation, the standard approach used in the quantitative
study of the spread of infectious diseases
\cite{brauer2012mathematical}. The basic idea of this kind of model is
to divide the relevant population into a set of classes or
compartments, which are defined according to the characteristics of
the disease to be studied, and then to model the population flows
between these compartments by means of a set of equations describing
the dynamics of the system. The results of the analytical and
numerical studies are interpreted in epidemiological terms for the
purpose of predicting the evolution of infections and, given the case,
defining epidemic control mechanisms. As these models became popular,
it was clear that their mathematical structure transcended the field
of epidemiology and that they were suitable for describing other
phenomena in the social and human sciences. As a consequence,
compartmental models have been used to study, with notable success,
processes as varied as the acceptance of new scientific ideas
\cite{goffman1964generalization, bettencourt2006power}, language
competition\cite{abrams2003modelling}, the propagation of memes and
rumors in social networks \cite{jin2013proceedings}, or the dynamics
of addictions \cite{bissell2014compartmental}.

In line with this approach, we propose a simple compartmental model to
describe the spread of radical ideas in a society. While the mere
existence of radical views does not necessarily pose a real problem,
some extreme cases such as terrorist groups, dangerous sects or
political extremists are a cause for concern in today's societies. For
this reason, the problem of radicalization has been extensively
addressed by the social sciences and, more recently, from
interdisciplinary perspectives including mathematical modeling
\cite{stares2007terrorism,galamplos,mccluskey2018bare,short2017modelling,nizamani2014public,banks2003bioterrorism}. In
this work, we rely on the analogy between, on the one hand, the spread
of an epidemic whose contagion occurs by direct contact between
infected individuals and susceptible individuals, and on the other
hand, the processes of radicalization considered in a simplified way
as the result of social interaction between individuals who already
profess and proselytize the radical ideology (the "infected"), and
individuals who are susceptible to adopt such ideology.

Compartmental models are usually based on a Markovian approximation
where the future evolution of the system depends only on the state of
the system at a single point in time. In many social systems, this
hypothesis is adequate as a first approximation; however, we cannot
ignore that, in these phenomena, memory (i.e. the history of the
system in a certain time interval) plays an important role. A natural
way to introduce memory into modeling is by using fractional calculus
\cite{hilfer2000applications}. Unlike ordinary derivatives,
derivatives of non-integer order also depend on the previous evolution
of the system, which makes a strong case for modeling systems with
memory. Since this was first observed, fractional models have been
used progressively in the study of this type of systems, reaching in
many cases a level of accuracy in the description much higher than
models based on standard calculus
\cite{sun2018new,duarte2019fractional}. In this article, we consider
the effects of the memory in our compartmental model of the
propagation of radical ideas in societies.

In the next section we present the hypotheses of the model and arrive
at the system of equations that governs its dynamics under the
Markovian approximation. After that, in Section \textcolor{black}{3} we
present the analytical and numerical study, obtaining and classifying
the equilibrium points, qualitatively showing the possible evolutions
of the system, and interpreting the results. In Section
\textcolor{black}{4}, we introduce the standard procedure that allows
the inclusion of memory in dynamical systems by means of Caputo's
derivatives.  \textcolor{black}{These ideas are applied to our model in
  Section \ref{sec:mem} where memory effects in the radicalization
  process are discussed. Finally, Section \textcolor{black}{5} is
  devoted to the conclusion. }

\section{The model}

Conceptualizing radicalization is a difficult task, on which there is
no consensus among experts. There is some agreement that it is a
gradual process, a product of socialization, by which individuals or
groups adopt an extremist worldview not shared by the mainstream of
society, and tend to justify the use of radical mechanisms (such as
the use of violence) to produce the desired social change
\cite{hafez2015radicalization}. Experts also differentiate between
\textit{cognitive radicalization}, which occurs at an ideological
level and whose characteristics are those described above, and
\textit{behavioral radicalization}, which can range from moderate,
public and legal activism in favor of the cause (for example, in
social networks), to explicit violent action
\cite{hafez2015radicalization}. These extreme cases constitute what is
called \textit{violent radicalization}, a process that generally
includes the use of specific mechanisms such as selective recruitment
and training. These characteristics, added to the fact that the number
of individuals involved in such actions is usually marginal, make it
difficult or directly question the relevance of a quantitative
analysis of the phenomenon \cite{kurzman2018missing}. On the other
hand, studies suggest that cognitive and moderate behavioral
radicalization are often a necessary condition for the emergence of
social violence \cite{horgan2009walking}, which is why we will focus
on understanding these processes.

It is therefore reasonable to propose compartmental models for the
study of radicalization. \textcolor{black}{Adapting} the terminology of
the SIR model, the population \textit{susceptible} to embrace the
radical idea is termed $S$, the population affiliated to that
\textcolor{black}{\textit{radical}} ideology and seeking to disseminate
it is termed \textcolor{black}{$R$, and $I$ the \textit{immunized}
  individuals who either left the radical group thus acquiring
  immunity}, at least transitory, or who are not very prone to adopt
that position (moderates, who may be present even in the absence of
radicals). Disregarding demographic changes due to births, deaths and
migratory flows, the sum of the populations of the three compartments
will be a constant, equal to the total population $N$:
\begin{equation}
S+\textcolor{black}{R+I}=N.
\end{equation}
Following the philosophy adopted in compartmental models, we establish
that all flows resulting from the interaction between individuals in
different compartments are assumed to be proportional to the product
of the populations. This assumption, usually adopted as a first
approximation, is known as the \textit{law of mass action} and implies
uniform mixing of the populations \cite{heesterbeek2005law}. In
addition, we will take all model parameters (spontaneous transition
rates, average number of contacts per unit time) as constants.

To model the dynamics of radical groups, we establish a set of
specific hypotheses about the phenomenon that exhibit clear
differences with traditional \textit{SIR} models:
\begin{itemize}
\item Radical individuals actively proselytize to convince those
  likely to join their group. The level of militancy of each radical
  individual is represented by a parameter $\alpha$. We assume that
  each radical has $\alpha$ effective contacts per unit time, $S/N$ of
  which occur with susceptible individuals. Thus, $\alpha
  \textcolor{black}{R} S /N$individuals will leave the group $S$ per
  unit of time.

\item The radical nature of the idea being propagated means that the
  above interactions may result not only in conviction, but also in
  active rejection. We will model this rejection by assuming that only
  a fraction $p$ of those who left the susceptible group by
  interaction with the radicals adhere to the group
  \textcolor{black}{$R$}, while the rest, by opposition, pass directly
  to the compartment \textcolor{black}{ $I$}, without having been
  \textcolor{black}{radicalized}. The parameter $p$ will depend on the
  efficiency of the social strategies employed, as well as on cultural
  factors.

\item The \textcolor{black}{immunized} population can intervene to
  prevent radicalization (or in the absence of radicalization, to
  combat high levels of susceptibility), which we will model as an
  interaction with susceptible individuals at a rate $\beta$ of
  effective contacts per \textcolor{black}{immunized} individual, per
  unit of time (in this case we will assume that the effect of this
  intervention can only be null or positive). Consequently, $\beta S
  \textcolor{black}{I}/N$ individuals per unit time will transit from
  the compartment $S$ to the compartment \textcolor{black}{$I$}. The
  parameter $\beta$ will be an indicator of the intensity of the
  actions on the vulnerable population, and, more abstractly, of the
  levels of activism and social solidarity.

\item Since radicalization typically arises out of opposition to a
  \textit{status quo} that one wishes to combat
  \cite{sedgwick2007inspiration}, we will assume that the interaction
  between \textcolor{black}{radicalized and immunized} individuals has a
  negligible effect in terms of the flow between the compartments. On
  the contrary, it is often observed that interventions seeking to
  disrupt radicalization lead to a strengthening of the sense of
  belonging among the members of the radical group
  \cite{tsintsadze2014groupthink}. In more extreme cases, the
  interaction with \textcolor{black}{immunized} individuals may be
  prevented for security reasons (as in the case of terrorist groups)
  or even outright prohibited (as in the case of some dangerous
  sects)\cite{hewstone2001social,tsintsadze2014groupthink,della2006social,lalich2004bounded}.

\item  Different reasons, loss of faith in the ideology, rejection of the strategies employed, desire to return to a normal life, can produce the decision to spontaneously leave the radical group. So we assume a constant recovery rate 
$\gamma$ which depends on the level of contentment provided by belonging to it, as well as on the group’s ability to retain its members. This will produce a $\gamma I$ flow of individuals  per unit time that will pass from compartment \textcolor{black}{ $R$} to compartment \textcolor{black}{$I$}.

\item Similarly, \textcolor{black}{immunized} individuals may, under
  certain conditions, become susceptible to radicalization. We will
  also assume that this occurs proportionally to the
  \textcolor{black}{immunized} population at a rate $\delta$, which,
  like $\gamma$, is the inverse of the average time of belonging to
  the group. According to this hypothesis, individuals per unit of
  time will move from compartment \textcolor{black}{$I$} to compartment
  $S$. Except for personal factors, $\delta$ may be linked to the
  level of social welfare and may increase in situations of economic
  or social crisis.
\end{itemize}

\begin{figure}  \centering
\includegraphics[width=0.7\columnwidth]{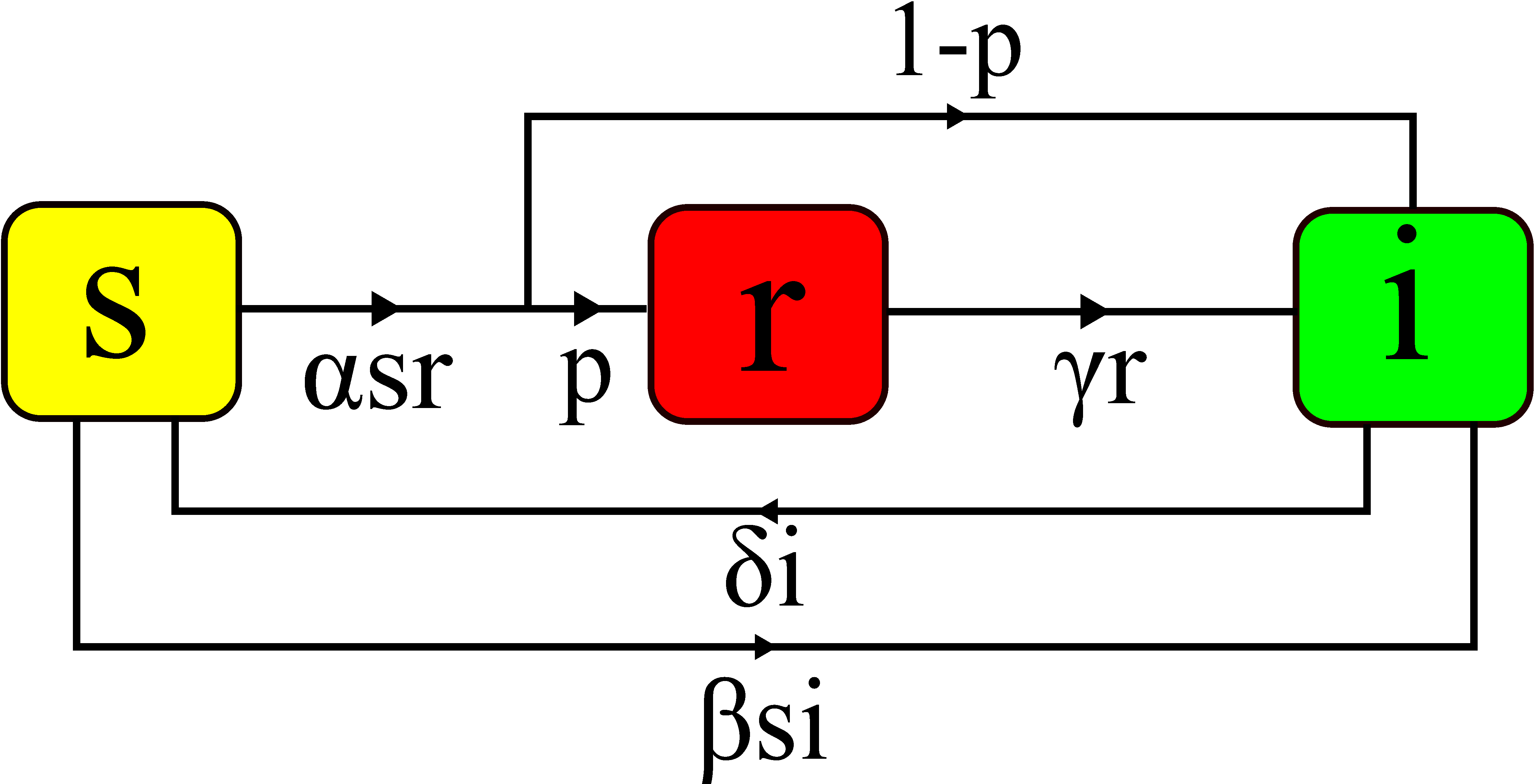} 
\caption{A schematic representation of the interactions between the
  populations.}
\label{figflow}
\end{figure}

Under the Markovian approximation, the above hypotheses result in the
following system of differential equations:
\begin{equation}\label{sistema}
\begin{cases}
\vspace{0.2cm}
\dfrac{ds}{dt}=-\alpha sr-\beta si+\delta i\\
\vspace{0.2cm}
\dfrac{dr}{dt}=p\alpha sr-\gamma r\\
\dfrac{di}{dt}=\gamma r+\beta si-\delta i+(1-p)\alpha sr
\end{cases}
\end{equation}
where $s$, $r$ and $i$ and denote the fractions of the total
population in each of the compartments at the point in time $t$ and
verify $s+r+i=1$. An schematic representation of the interactions
between the population is given in Fig.~\ref{figflow}.

\section{Analysis in the Markovian model}
\subsection{Equilibrium points}
From Eq.\ref{sistema} we easily obtain the equilibrium points of the system:
\begin{equation}
\begin{split}
&P_{1}=\left(1,0,0\right),\\&P_{2}=\left(\dfrac{\delta}{\beta},0,1-\dfrac{\delta}{\beta}\right),\\
&P_{3}=\left(\dfrac{\gamma}{p\alpha},\dfrac{(p\alpha\delta-\beta\gamma)(p\alpha -\gamma)}{p\gamma(p\alpha \delta-\beta\gamma+\alpha\gamma)},\dfrac{\gamma (p\alpha -\gamma)}{p(p\alpha\delta-\beta\gamma+\alpha\gamma)}\right).
\end{split}
\end{equation}
We verify that $P_{1}$ is an equilibrium point for any set of
parameter values (which corresponds to the \textit{disease-free
  equilibrium} of epidemiological
models\cite{brauer2012mathematical}). Given that $s,r,i\leq 1$,
$P_{2}$ is also an equilibrium point only if $\delta \leq\beta$. As
for $P_{3}$, we deduce the necessary condition of existence,
$\gamma\leq p\alpha$, by observing its first component (C1). Then,
from the non-negativity of the third component, we obtain
$p\alpha\delta-\beta\gamma+\alpha\gamma>0$ (C2). This result, together
with the non-negativity of the second component leads to the
additional condition $p\alpha\delta-\beta\gamma>0$ (C3). Given that
$\alpha\gamma\geq 0$, it is evident that condition (C3) implies
(C2). In summary:
\begin{equation}\label{existencia1}
\begin{split}
\vspace{0.2cm}
&\exists P_{1} \hspace{0.2cm}\forall \alpha,\beta,\gamma,\delta,p\\
\vspace{0.2cm}
&\exists P_{2}\leftrightarrow \delta\leq\beta\\
\vspace{0.2cm}
&\exists P_{3}\leftrightarrow \gamma\leq p\alpha\hspace{0.2cm}\wedge \hspace{0.2cm}p\alpha\delta\geq\beta\gamma
\end{split}
\end{equation} 
The analysis of the above conditions shows that it is possible to
significantly simplify the problem, reducing the number of parameters
from five to two by defining the following:
\begin{equation}
\Omega=\dfrac{\beta}{\delta};\hspace{0.2cm}\Lambda=\dfrac{p\alpha}{\gamma}.
\end{equation} 
In terms of the new parameters, the conditions of existence, 
Eq.~(\ref{existencia1}), are expressed as
\begin{equation}\label{existencia2}
\begin{split}
\vspace{0.2cm}
&\exists P_{1} \hspace{0.2cm}\forall\hspace{0.2cm} \Omega,\Lambda\\
\vspace{0.2cm}
&\exists P_{2}\leftrightarrow \Omega\geq 1\\
\vspace{0.2cm}
&\exists P_{3}\leftrightarrow \Lambda\geq 1 \hspace{0.2cm}\wedge \hspace{0.2cm}\Lambda\geq\Omega\leftrightarrow \Lambda\geq\text{max}(1,\Omega)\end{split}
\end{equation} 

Parameters $\Lambda$ and $\Omega$ fully characterize the long term
dynamics of the system. $\Lambda$ is the average number of susceptible
individuals recruited by each radicalized individual while remaining
in the group and we will refer to it as \textit{radical strength}
because it is a measure of the \textit{recruitment and retention}
capability of the radical group. This parameter is analogous to
the\textit{ basic reproduction number} in epidemiological models
\cite{heesterbeek2005law}.  Similarly, $\Omega$ represents the
expected value of the number of individuals that each
\textcolor{black}{immunized} individual managed to remove from the
susceptible state. A society will present a high value of $\Omega$ if,
in parallel to reasonable levels of welfare that decrease the
probability of becoming susceptible (small $\delta$), there is an
important level of positive interaction among its members (large
$\beta$), so it can be considered an indicator of \textit{social
  cohesion} \cite{kawachi2000social,alcala2017social}.

Regarding the initial states that we will consider, in the absence of
radical groups the dynamics described by Eqs.~\ref{sistema} is
equivalent to that of the model \textit{SIS} for the spread of
diseases that do not provide immunity
\cite{brauer2012mathematical}. The initial state considered will
consist of a small fraction of radicalized individuals for the purpose
of studying their evolution.

\subsection{Stability analysis and interpretation of results}
From the conditions (\ref{existencia2}) we can show that there are
five regions in the parameter space $\left(\Lambda,\Omega\right)$ (see
Fig.~\ref{fig:f7} below), characterized by the existence of the
different possible combinations of equilibrium points. In order to
simplify the stability study of the system (\ref{sistema}), we use the
conservation of the total population and take as variables the
\textcolor{black}{radicalized and the immunized,} resulting in the
following:
\begin{equation}\label{sistrucho}
\begin{cases}
\begin{split}
&\dfrac{dr}{dt}=(p\alpha-\gamma)r-p\alpha ri-p\alpha r^{2}\\
&\dfrac{di}{dt}=(\gamma +(1-p)\alpha)r+(\beta-\delta)i-\beta i^{2}\\&\hspace{1.0cm}-(\beta+(1-p)\alpha)ri-(1-p)\alpha r^{2}.
\end{split}
\end{cases}
\end{equation}
from which the Jacobian matrix can be easily obtained.

\subsubsection{Case 1: $\Omega <1 \hspace{0.2cm}\wedge\hspace{0.2cm} \Lambda <1$}
Given that $\Omega <1$, society is in the susceptible state prior to
the appearance of the radical group the, and according to the
discussion (\ref{existencia2}), under these conditions the only
equilibrium point of the system (\ref{sistema}) is $P_{1} = \left(
1,0,0 \right)$. The eigenvalues of the Jacobian matrix at that point
are the following:
\begin{equation}\label{eigP1}
\begin{cases}
\lambda^{P_{1}}_{1}=p\alpha -\gamma =\gamma (\Lambda -1)<0\\
\lambda^{P_{1}}_{2}=\beta -\delta =\delta (\Omega -1)<0
\end{cases}
\end{equation}
Therefore, it will be asymptotically stable. Should a small radical
group appear (regardless of size), it will disintegrate and the
population will quickly return to the susceptible state.

\begin{figure}[h]
 \centering
 \includegraphics[width=0.65\columnwidth]{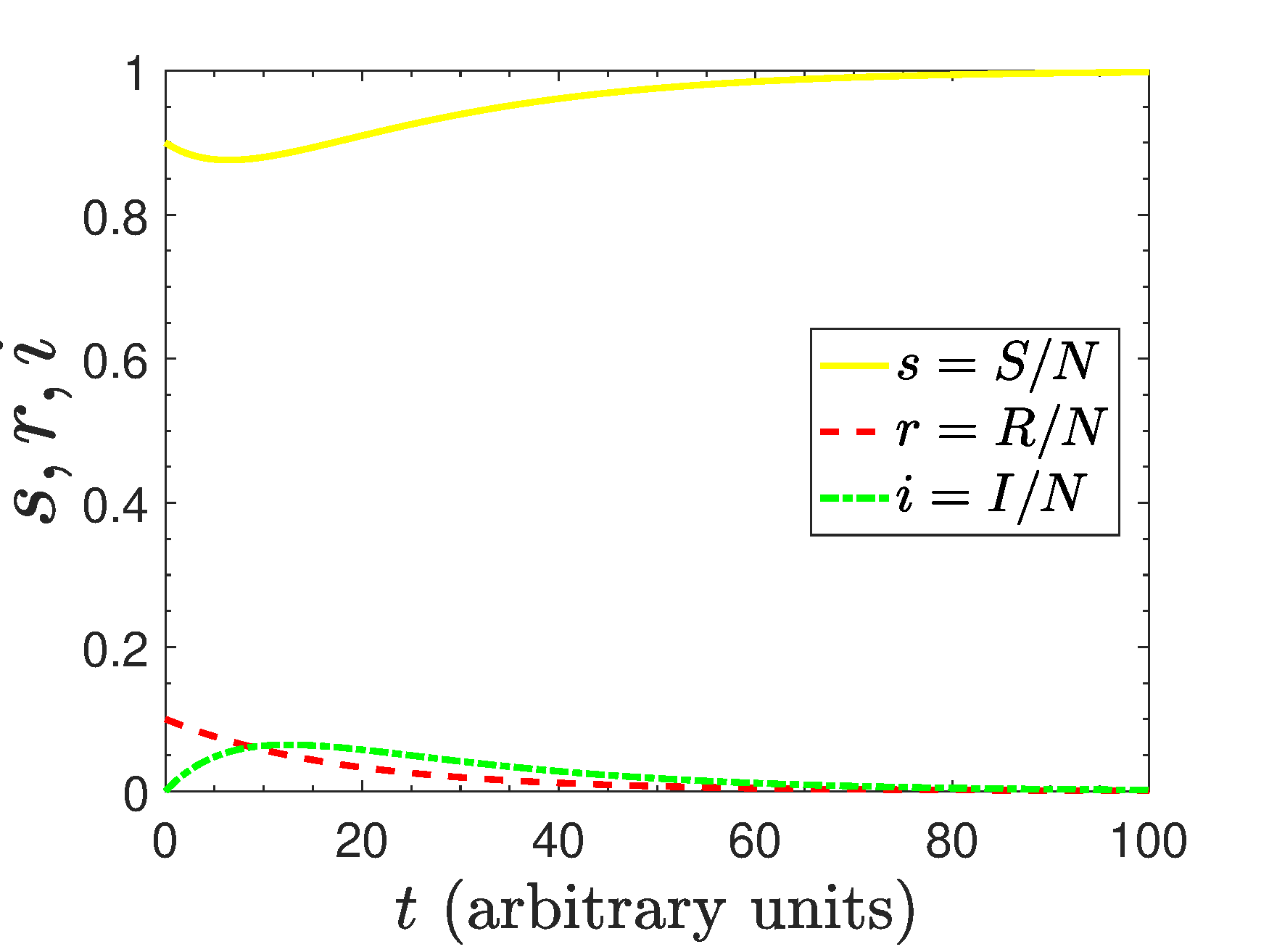} \caption{
   Evolution of the populations in case 1: in a society with low
   social cohesion and poor capacity of the radical groups for
   attracting individuals the population radicalized shows an
   exponential decay. Initial distribution $(0.9,0.1,0)$ and parameter
   values: $\alpha=0.1$, $\beta = 0.1$, $\gamma = 0.1$, $\delta=0.2$
   and $p=0.5$ ($\Omega =\Lambda =1/2<1$).}
\label{f:f2}
\end{figure}

This situation corresponds to societies with low social cohesion
($\Omega <1$), but in which the radical group does not manage to
develop a good capacity for attracting susceptible individuals or
retaining its current members ($\Lambda<1$). Figure~\ref{f:f2} shows
the temporal evolution of the populations of the three compartments
for a somewhat artificial case in which 10\% of the population becomes
radicalized, showing an initial exponential decay in this group. This
can be understood by noting that the equation governing that
compartment at the beginning of the evolution is, approximately:
\begin{equation}\label{inicialinfected}
\dfrac{dr}{dt}=(p\alpha s_{0}-\gamma)r=\gamma(\Lambda s_{0}-1)r
\end{equation}
and under the condition $\Lambda <1$, the first factor is negative for
any initial fraction of susceptible individuals $s_{0} <1$. That decay
produces a flow towards \textcolor{black}{$I$}, a group that is quickly
abandoned due to the low levels of cohesion.

\subsubsection{Case 2: $\Omega >1 \hspace{0.2cm}\wedge\hspace{0.2cm} \Lambda <1$}

In this case both $P_{1}$ and $P_{2}$ are equilibrium points. The
condition $\Omega >1$ implies that $\lambda^{P_{1}}_{2}>0$, therefore
$P_{1}$ now has an unstable direction (saddle point). As for $P_{2}$,
the eigenvalues at that point are the following:
\begin{equation}\label{eigP2}
\begin{cases}
\lambda^{P_{2}}_{1}=\dfrac{p\alpha\delta -\beta\gamma}{\beta}
=\gamma\left(\dfrac{\Lambda}{\Omega}-1\right)
<0\\ \lambda^{P_{2}}_{2}=\delta -\beta =\delta(1-\Omega)<0
\end{cases}
\end{equation}

\begin{figure}[h]
\centering
\includegraphics[width=0.65\columnwidth]{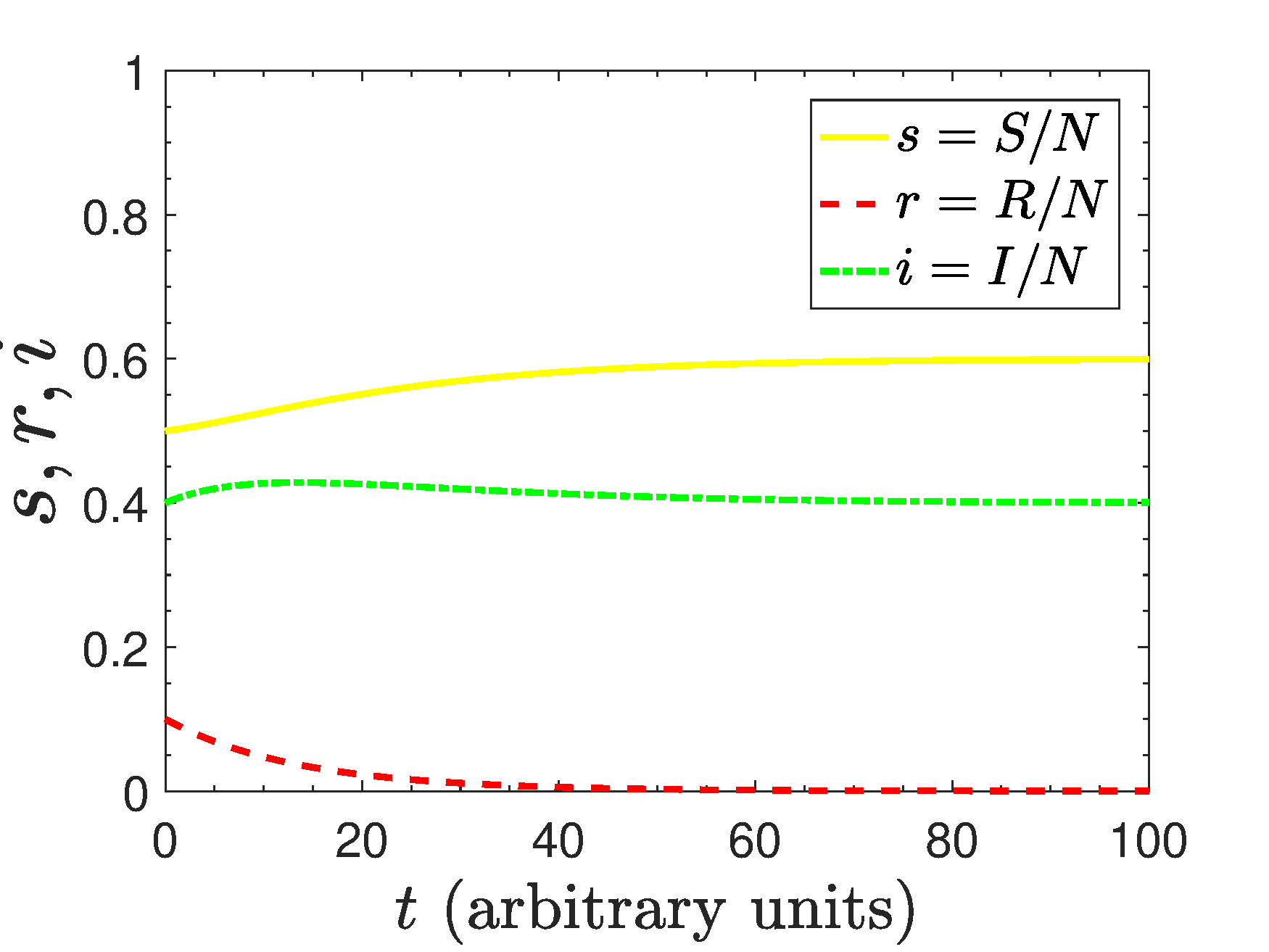}
\caption{Coexistence of susceptible and moderate groups in a society
  with high cohesion levels and radical groups exhibiting weak
  strength. The initial condition is $(0.5, 0.1, 0.4)$ and the
  parameter values are $\alpha=0.1$, $\beta = 1/6$, $\gamma = 0.1$,
  $\delta=0.1$ and $p=0.5$ ($\Omega =5/3>1$, $\Lambda =1/2<1$).}
\label{f:f3}
\end{figure}

Note that if we start from case 1 and we increase $\Omega$, a forward
bifurcation occurs in $\Omega=1$ \cite{gumel2012causes} so $P_{1}$
loses its stability as the new stable equilibrium point is
established:
\begin{equation}
P_{2}=\left(\dfrac{1}{\Omega},0,1-\dfrac{1}{\Omega}\right).
\end{equation}
This equilibrium point is the one that exists before the emergence of
the radical group and is stable in both cases. This implies that the
inclusion of some radicals will not modify the distribution, and the
population will continue to be distributed between the susceptible and
\textcolor{black}{immunized} categories in the same proportions as
before, defined by $\Omega$. This situation corresponds to the case of
societies with high cohesion levels ($\Omega >1$), in which the
strength of the radical group remains at low levels ($\Lambda <1$). In
particular, in the limit, $\Omega\gg 1$ the majority of the society
will belong to the moderate group. The evolution of the populations
for the same initial situation as in the previous case is shown in
Fig.~\ref{f:f3}, where, the yellow and green curves will be
approximately constant, given that $\textcolor{black}{r}(0)\ll 1$.

\subsubsection{Case 3: $\Omega <1 \hspace{0.2cm}\wedge\hspace{0.2cm} \Lambda >1$}
As noted, the condition $\Omega <1$ implies that in the absence of
radicals, the entire population will be susceptible. By introducing a
small fraction of radicals, the dynamics becomes governed by the
system (\ref{sistema}), which will have $P_{1}$ and $P_{3}$ as
equilibrium points. From Eq.~\ref{eigP1} we observe that
$\lambda^{P_{1}}_{1}>0$ and $\lambda^{P_{1}}_{2}<0$, therefore $P_{1}$
is a saddle point. The proof of the asymptotic stability of $P_{3}$,
which is cumbersome due to the complex expression of the eigenvalues,
can be analyzed with the signs of the trace and the determinant of the
Jacobian matrix. Again, if we start from case 1 and increase $\Lambda$
a \textit{forward bifurcation} occurs in $\Lambda =1$.

\begin{figure}[h]
\centering
\includegraphics[width=0.65\columnwidth]{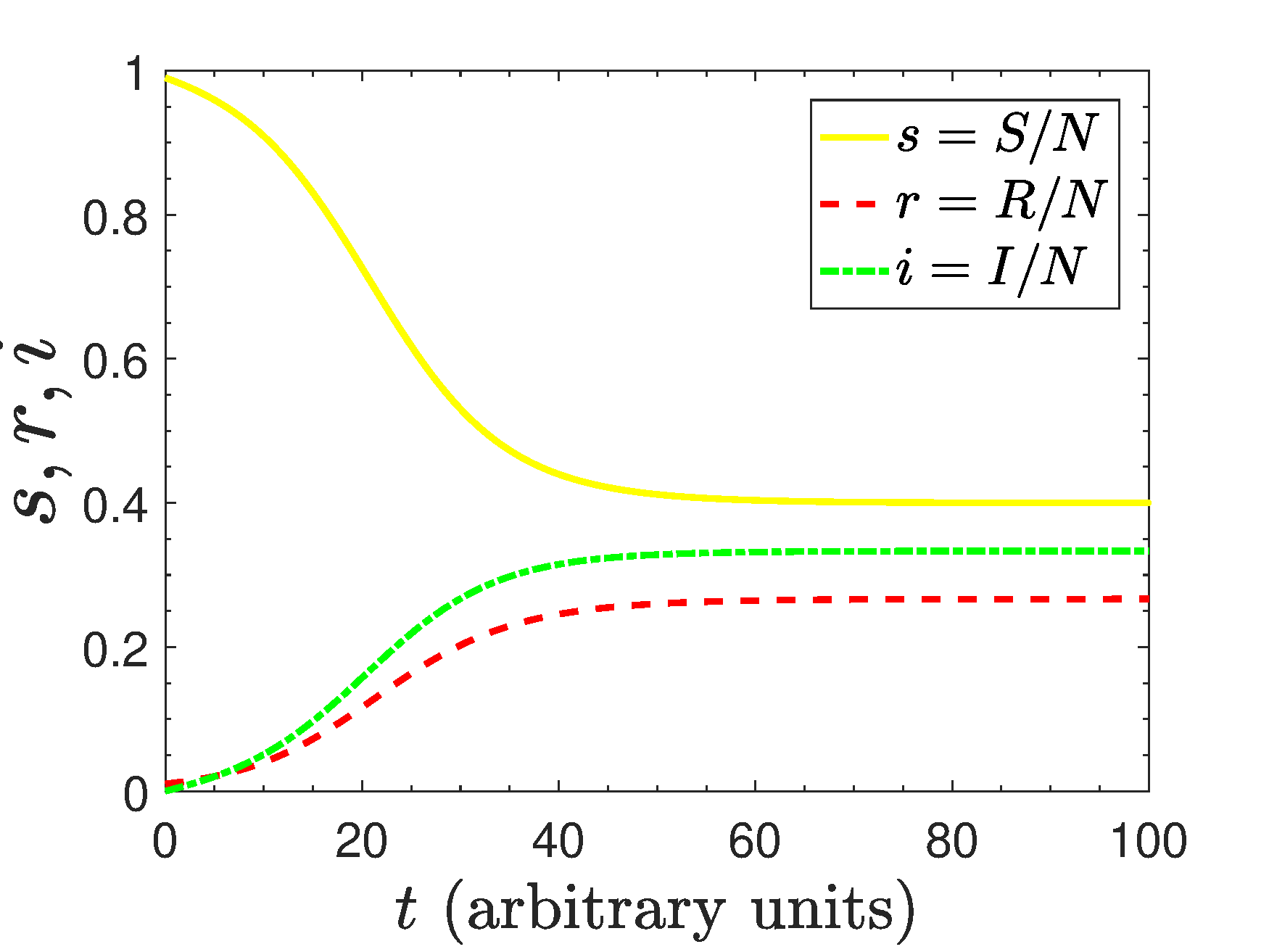}
\caption{Strong radical groups in societies with low cohesion levels
correspond to case 3. The values taken for the parameters are $\alpha=0.5$, $\beta = 0.1$, $\gamma = 0.1$, $\delta=0.2$ and $p=0.5$ ($\Omega =1/2<1$, $\Lambda =5/2>1$). The initial distribution of populations is $(0.99,0.01,0)$.}
\label{f:f4}
\end{figure}

This case corresponds to radical groups that evolved towards more
efficient organizations in at least one of the aspects of recruitment
and retention and that are immersed in weakly cohesive
societies. Consequently, the dynamics, shown in Fig.~\ref{f:f4},
reveals a sustained growth of such a group, whose proportion will
constitute a non-negligible fraction of the total population in the
asymptotic regime. In fact, in the $\Lambda \gg 1$ limit it is
possible that large percentages of the population will end up adhering
to the radical group. There are plenty of historical examples where
this has happened, for example at the political level as a consequence
of the emergence of charismatic radical leaders in societies
undergoing serious periods of crisis \cite{pappas2008political}.

We also note that despite the high social fragility ($\Omega <1$), the
asymptotic state can present an considerable number of moderates, a
fact that did not occur prior to the appearance of the radical
group. It is reasonable to suppose then that this occurs mainly as a
response to radicalization, linked to the fraction $(1-p)$ and not so
much due to the abandonment of the radical group nor to the
intervention capacity of the \textcolor{black}{immunized} individuals.

\subsubsection{Case 4: $\Omega >\Lambda>1$}

\begin{figure} \centering
\includegraphics[width=0.65\columnwidth]{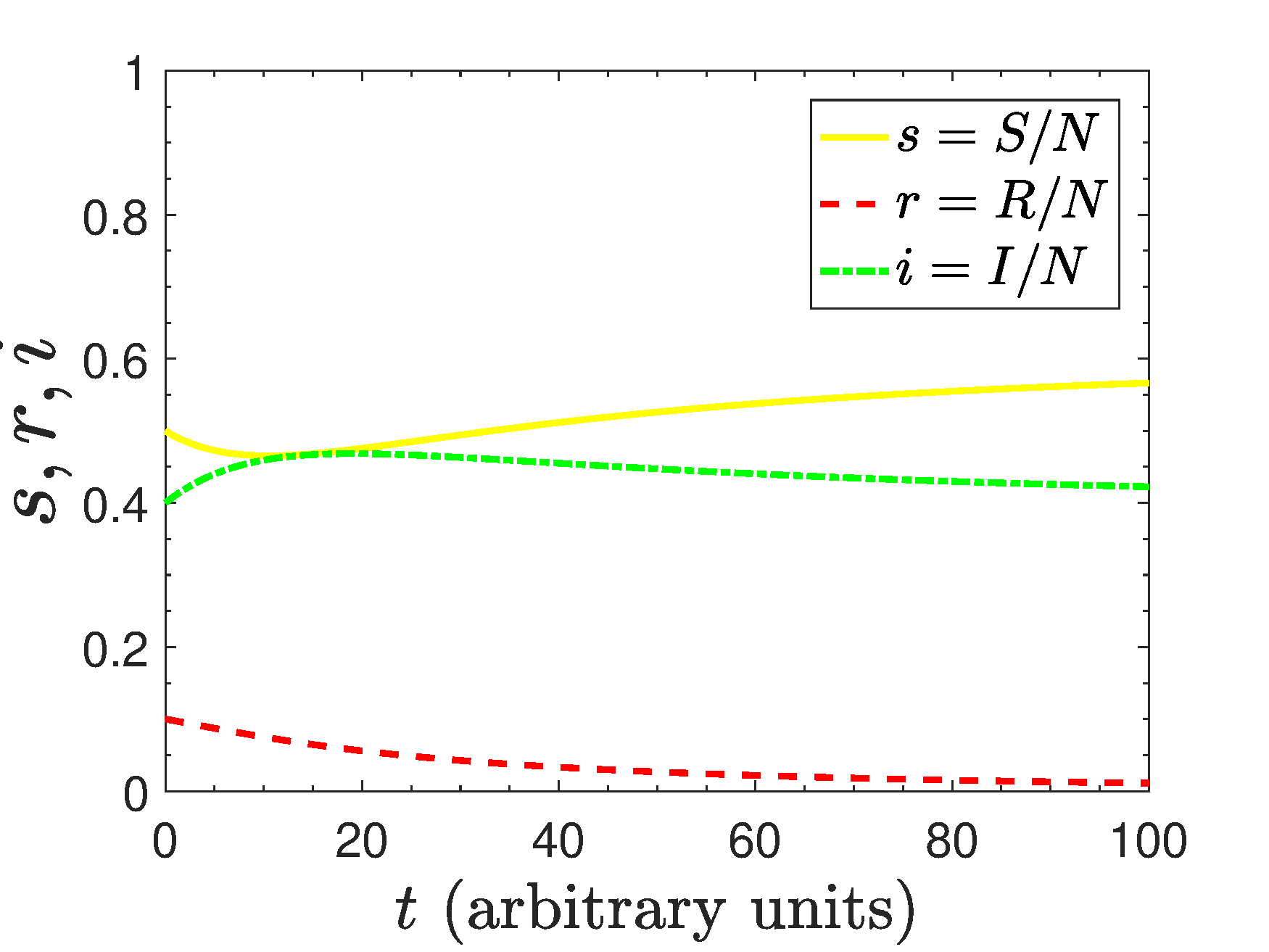}
\caption{In societies with high cohesion levels the radical groups
  will remain in a marginal level regardless of their strength in case
  4. The values of the parameters are $\alpha=0.3$, $\beta = 1/6$,
  $\gamma = 0.1$, $\delta=0.1$ and $p=0.5$ ($\Omega =5/3>\Lambda
  =3/2>1$). The initial distribution of populations is
  $(0.5,0.1,0.4)$.}
\label{f:f5}
\end{figure}

Under these conditions the only equilibrium points are $P_{1}$ and
$P_{2}$. The long-term behavior is identical to that of case 2, since
the only difference is that $P_{1}$ goes from being a saddle point to
being unstable. Therefore, the equilibrium state coincides again with
the state prior to the appearance of the radicals, whose long-term
expression is again marginal. This occurs despite the fact that the
radicals have increased their strength, a fact that manifests itself
in two ways: on the one hand, in a longer dispersion time of that
group; on the other hand, in a scenario of light growth at the
beginning of the evolution, given that, if $\Lambda$ is large enough,
$\Lambda s_{0}$ can be greater than $1$ (see
Eq.~\ref{inicialinfected}). This does not occur for the parameter
values chosen in Fig.~\ref{f:f5}, but it was verified for other
values.  This case shows that in sufficiently cohesive societies,
radical ideas will have little place regardless of their strength. As
we will see below, this fact can be used to develop containment
strategies.

\begin{figure}[h]
 \centering
    \includegraphics[width=0.65\columnwidth]{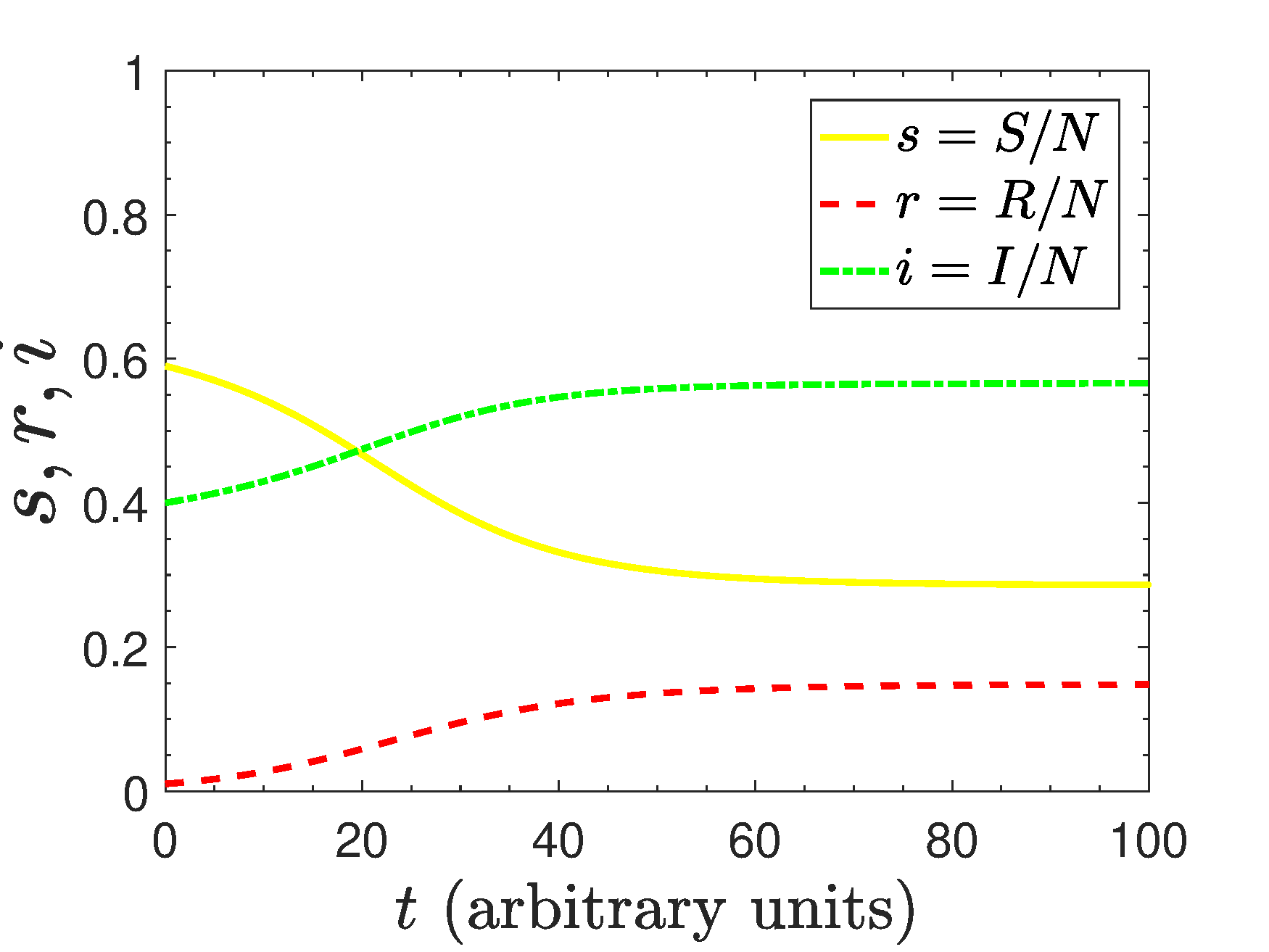}
   \caption{Case 5 corresponds to highly polarized societies.  The
     values of the parameters are $\alpha=0.7$, $\beta = 1/6$, $\gamma
     = 0.1$, $\delta=0.1$ and $p=0.5$ ($\Lambda =7/2>\Omega
     =5/3>1$). The initial distribution of populations is
     $(0.59,0.01,0.4)$}
\label{f:f6}
\end{figure}

\subsubsection{Case 5): $\Lambda >\Omega >1$}
We note, once again, that the coexistence of three equilibrium points,
$P_{1}$ (unstable), $P_{2}$ (saddle point) and $P_{3}$ (asymptotically
stable), implies that radicals will have an important
expression. However, faced with a larger set of moderates than in case
3, since the higher social cohesion ensures the functioning of the
other \textcolor{black}{immunization} mechanisms. It is also possible to
see that as increases $\Lambda$ for a $\Omega$ fixed, the number of
susceptible individuals in the equilibrium state decreases, resulting
in a polarized society between radical and \textcolor{black}{immunized}
individuals. The evolution of the populations is shown in
Fig.~\ref{f:f6}.

\begin{figure}
 \centering
 \includegraphics[width=.6\columnwidth]{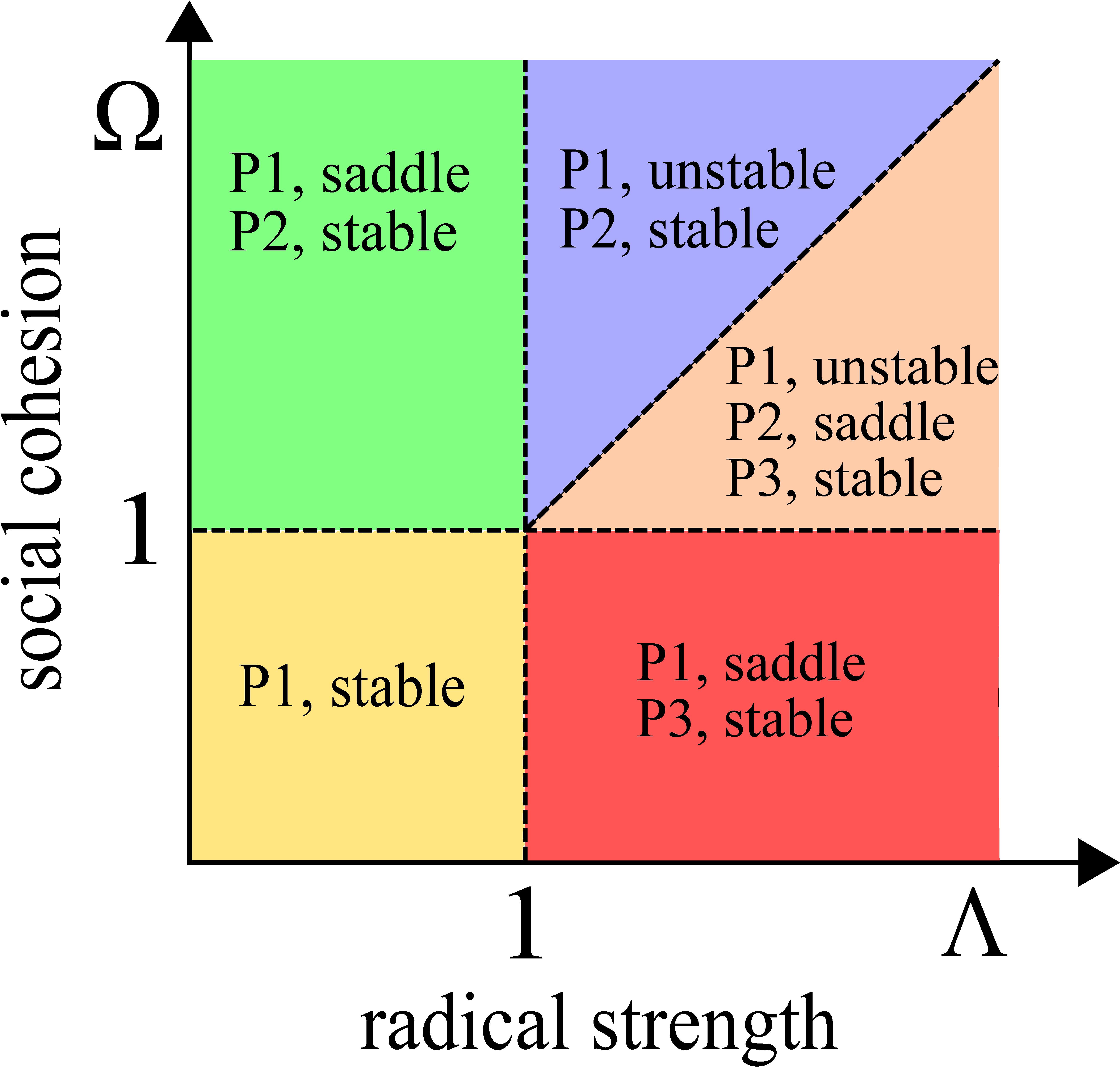}
 \caption{Relevant regions in the parameter space ($\Lambda$,
   $\Omega$), defined as a function of the present equilibrium points
   and their stability.}\label{fig:f7}
\end{figure}

\section{Modeling of systems with power-law memory: the Caputo derivative}

It is important to note that when modeling some aspect of human
societies using ordinary differential equations (such as
Eq. \ref{sistema}), we are not taking the effects of memory into
account.  \textcolor{black}{Although this a reasonable approximation} in
many cases of interest, it is to be expected that in situations such
as the one studied in this work, the history of individuals and of
society itself will play a non-negligible role.

\textcolor{black}{Specifically,} there is a broad evidence that the
processes that involve the use of memory in humans follow power laws
\cite{edelman2020evolution}. In particular, various experiments have
shown that the decrease in efficiency achieved in memorization-based
tasks decreases as $ t^{-\epsilon}$, where $0 <\epsilon <1 $
\cite{kahana2012foundations}.  Reciprocally, learning (measured in
terms of the decrease in reaction times to certain stimuli) also
follows a power law in the number of attempts
\cite{anderson2000learning}. Usually this behavior of memory in
mammals is explained as a consequence of the corresponding
\textit{power-law memory} that neurons present, as well as a good part
of the rest of cells and tissues due to to their viscoelastic
character.

\textcolor{black}{ As we have already commented, the inclusion of memory
  in the evolution equations can be implemented by using fractional
  calculus.} To introduce the fractional derivatives, we start from
the Cauchy formula for iterated integrals of order $n\in N$:
\begin{equation}
I^{n}_{0}f=\dfrac{1}{(n-1)!}\int_{0}^{t}\left(t-t'\right)^{n-1}f(t')dt'
\end{equation}
and extend it to an arbitrary positive index $b$, using Euler’s 
$\Gamma$ function as an
analytic continuation of the factorial over the real numbers. Thus, we obtain
the \textit{Riemann-Liouville fractional integral}:
\begin{equation}\label{riemmann}
I^{b}_{0}f=\dfrac{1}{\Gamma(b)}\int_{0}^{t}\left(t-t'\right)^{b-1}f(t')dt'.
\end{equation}

Among the different definitions of fractional derivatives, in this
work we will rely on the Caputo derivative, which, for an arbitrary
order of differentiation $a>0$, is defined as \cite{caputo1967linear}
\begin{equation}\label{Caputo}
^{C}D^{a}_{0}f=I^{n-a}D^{n}f=\dfrac{1}{\Gamma(n-a)}\int_{0}^{t}\left(t-t'\right)^{n-a-1}D^{n}f(t')dt',
\end{equation}
where $n$ is the first integer greater than $a$, and $D^{n}$ is the
usual differentiation operator of order $n$.

\textcolor{black}{
The first advantage that the use of this derivative presents with 
respect to other definitions is
that initial value problems with Caputo derivative are initialized by 
integer-order derivatives \cite{garrappa2018numerical}:
\begin{equation}
\begin{cases}
^{C}D^{a}_{0}x(t)=f(t,x)\\
x(0)=x_{0},\hspace{0.1cm}x'(0)=x'_{0},...,x^{(n-1)}(0)=x^{n-1}_0,
\end{cases}
\end{equation}
unlike what occurs when employing other definitions (such as the
Riemann-Liouville derivative, which requires the knowledge of
fractional derivatives as initial values).}

\textcolor{black}{ Another important advantage of the Caputo derivative
  is related to its action on constant functions. Since the
  derivatives of any order of a constant function $f$ are zero for all
  times, Eq. \eqref{Caputo} implies that
\begin{equation}
^{C}D^{a}_{0}f=0,
\end{equation}
just like in the standard case.  As a consequence, if we consider a
system of ordinary differential equations (such as
Eq. \eqref{sistema}), and substitute all derivatives by fractional
derivatives, the equilibrium points of the fractional system will be
the same as those of the original system. It is important to emphasize
that this does not occur with other definitions of fractional
derivative \cite{garrappa2018numerical}.}  garr \textcolor{black}{ In
  order to understand how the Caputo derivative can play an important
  role in the modeling of systems with memory, let us consider a first
  order differential equation of the type:
\begin{equation}\label{eq_general}
D^{1}x(t)=f(x(t))
\end{equation}
where $D^{1}$ is the common derivative. Since the rate of change of
the unknown function $x(t)$ depends only on the value of $f$ at the
present time, the system described by Eq. \eqref{eq_general} has no
memory. To include memory in the system, we will assume that the rate
of change $D^{1}x(t)$ is no longer equal to the value of $f$ at that
point, but rather depends on the previous values of $f$ through its
convolution with a function $\varphi$, called the \textit{memory
  kernel}:
\begin{equation}\label{convolution}
D^{1}x(t)=\int_{0}^t\varphi(t-t')f(x(t'))dt'
\end{equation}
}

Different choices of $\varphi$ are associated with different types of
memory. For example, if we take as kernel the \textit{Dirac delta}
distribution, we obtain Eq.~(\ref{eq_general}), which corresponds to
the Markovian case. On the opposite hand, choosing the derivative of
the previous distribution as kernel, known as the \textit{Heaviside
  step function}, implies that the system has infinite memory, since
it assigns equal weight to all the previous points.

Since our interest is to model human societies, according to the
empirical evidence already discussed, it is reasonable to choose
$\varphi$ functions that decay as power laws. In particular
\begin{equation}
\varphi(t-t')=\dfrac{\left(t-t'\right)^{a-2}}{\Gamma (a-1)}.
\end{equation}
By replacing this kernel in Eq. (\ref{convolution}), we obtain the following: 
\begin{equation}\label{eq_general2}
D^{1}x(t)=\dfrac{1}{\Gamma(a-1)}\int_{0}^{t}\left(t-t'\right)^{a-2}f(x(t'))dt'.
\end{equation}
It should be noted, however, that by virtue of the definition
(\ref{riemmann}), the r.h.s. of Eq.~\ref{eq_general2} is nothing other
than the Riemmann-Liouville fractional integral of order $a-1$:
\begin{equation}
\dfrac{1}{\Gamma(a-1)}\int_{0}^{t}\left(t-t'\right)^{a-2}f(x(t'))dt'=I_{0}^{a-1}f(x(t)),
\end{equation}
therefore, Eq. (\ref{convolution}) becomes
\begin{equation}
D^{1}x(t)=I_{0}^{a-1}f(x(t)).
\end{equation}
Applying the  $I_{0}^{1-a}$ operator to both sides of the above equation,
we obtain the following
\begin{equation}
I_{0}^{1-a}D^{1}x(t)=I_{0}^{1-a}I_{0}^{a-1}f(x(t)).
\end{equation}

Finally, note that the composition of the operators on the right-side
term is equivalent to the identity operator, while the left-side term
is, by virtue of the definition \cite{caputo1967linear}, the Caputo
derivative of order $a$ ($0<a\leq 1$), so we obtain the following:
\begin{equation}
^{C}D^{a}_{0}x(t)=f(x(t)).
\end{equation}

In summary, if we assume that the rate of change of the function we
seek, $x(t)$, does not depend on $f$ at the point, but rather on $f$
weighed by a memory kernel that decays as a power law, then the
dynamics of the system will no longer be governed by an ordinary
differential equation, but by one of fractional order less than 1
(equivalent to an ordinary integral-differential equation). It is also
possible to see that the degree of memory of the system is controlled
by the order of the derivative: as approaches the value $1$, the
system has less memory \cite{du2013measuring}, and in the extreme case
of $a=1$ we return to the Markovian model, which is described
correctly by an ordinary differential equation.

Applying these ideas to the model described above, we obtain the
following fractional system of order $a$:
\begin{equation}\label{sistemaF}
\begin{cases}
\vspace{0.2cm}
^{C}D^{a}_{0}s=-\alpha sr-\beta si+\delta i\\
\vspace{0.2cm}
^{C}D^{a}_{0}r=p\alpha sr-\gamma r\\
^{C}D^{a}_{0}i=\gamma r+\beta si-\delta i+(1-p)\alpha sr
\end{cases}
\end{equation}which will be analyzed in the next subsection.

\section{Memory effects in the radicalization process}
\label{sec:mem}
As we have already mentioned, the equilibrium points of the system
(\ref{sistemaF}) coincide with those of the analyzed Markovian
case. The asymptotic stability in the fractional case is assured if
the two eigenvalues of the Jacobian matrix of the system, evaluated at
that point, satisfy the condition $\vert\text{Arg}(\lambda_{i})\vert >
a\pi /2$ \cite{matignon1996stability}.  This result elegantly
generalizes the well-known result for ordinary systems
\cite{strogatz2018nonlinear}. In particular, if the eigenvalues are
real, their argument can only be $0$ (positive real) or $\pi$
(negative real). Since we work under the restriction $0<a\leq 1$, and
by virtue of Eqs. (\ref{eigP1}) and (\ref{eigP2}), we conclude that
$P_{1}$ and $P_{2}$ (and hence also $P_{3}$) will be asymptotically
stable in the same regions as in the ordinary problem. This implies
that the inclusion of memory does not produce any modification
regarding the final distribution of the populations of the
system. However, we will see that it does have significant effects on
the evolution towards the asymptotic state.


\begin{figure}[!h] \centering
\includegraphics[width=0.49\columnwidth]{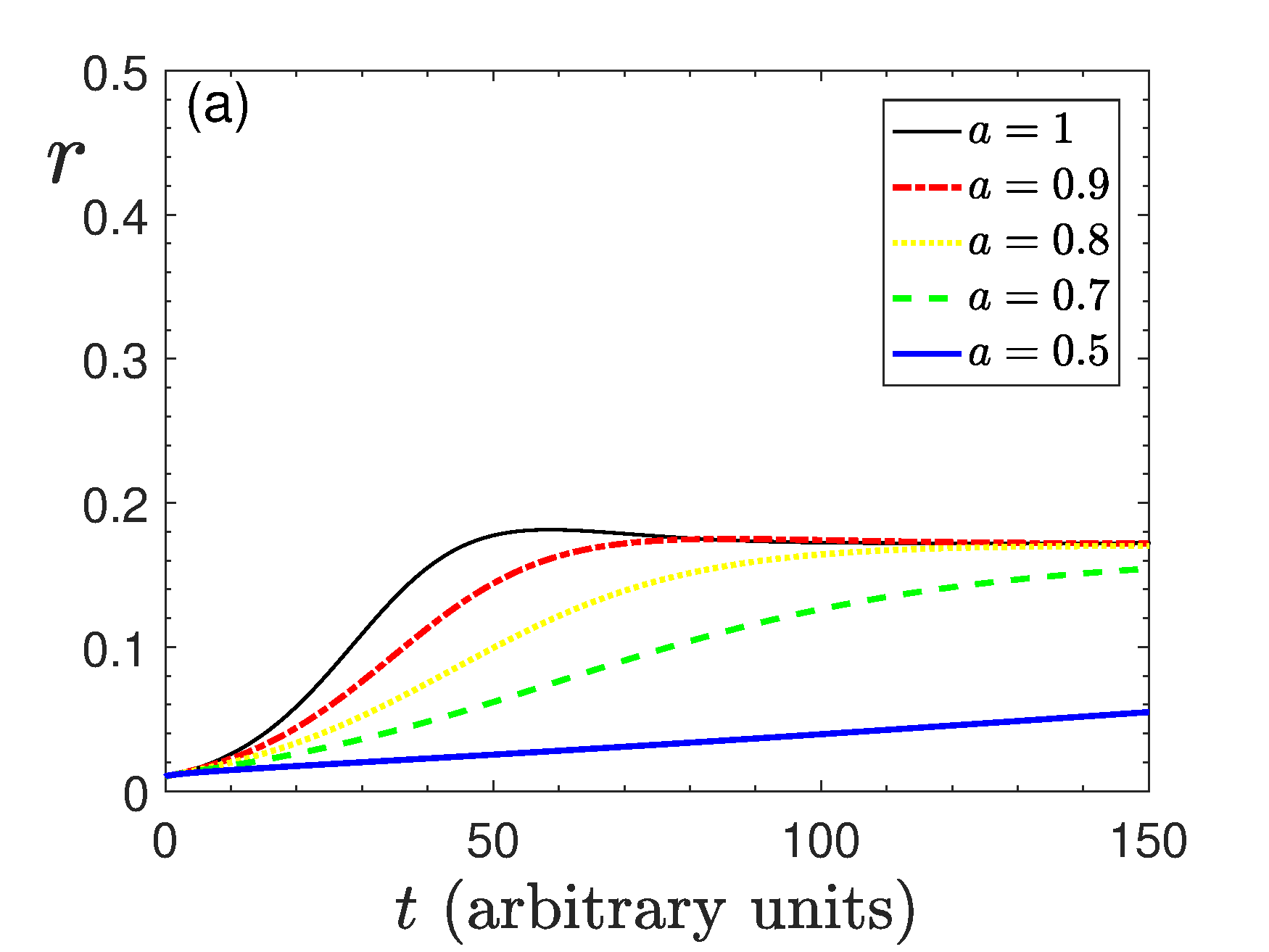}
\includegraphics[width=0.49\columnwidth]{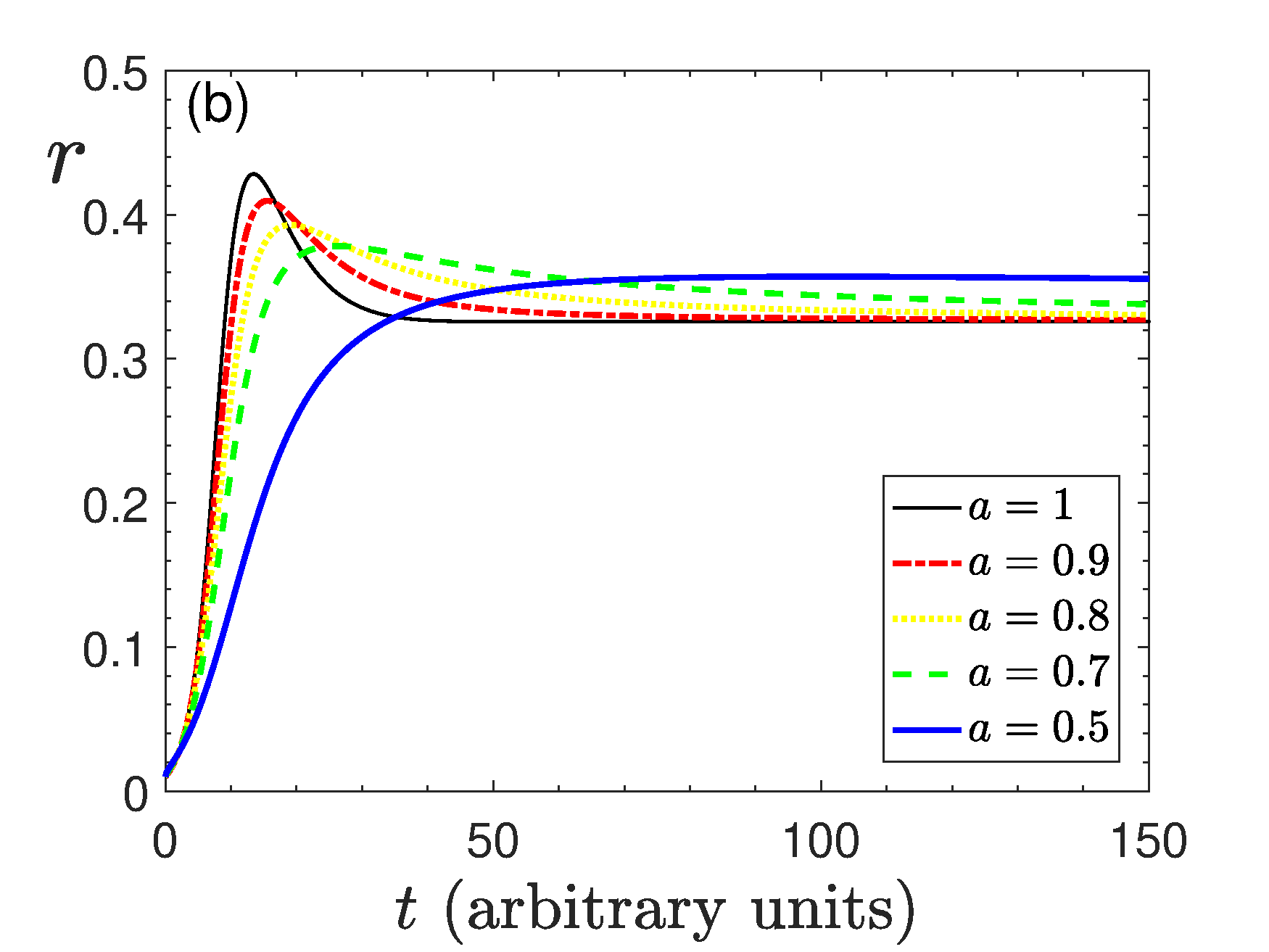}
\caption{Evolution of the \textcolor{black}{radicalized} for case 3,
  considering different levels of memory and different strengths of
  the radical group. Parameter values are such that: a) $\Omega =1/2$,
  $\Lambda =9/5$; b) $\Omega =1/2$, $\Lambda=6$. The numerical methods
  employed are described in Ref. \cite{garrappa2018numerical}. }
\label{f:f8}
\end{figure}

Figure \ref{f:f8} shows the evolution of the
\textcolor{black}{radicalized} individuals for two situations
corresponding to case 3, but at different limits of strength of the
radical group. In the first graph (relatively strong radical group) we
observe that the Markovian approximation underestimates the duration
of the transient regime, the longer the times required to reach the
steady state the more memory is included in the system (the time
scales required can be very different). On the other hand, the second
graph shows that, in the presence of a very strong radical group,
models with little memory predict the existence of radicalization
peaks before the final distribution is reached. When more memory is
included, this behavior tends to disappear and in the infinite memory
limit the radicals grow monotonically until reaching their equilibrium
value. Although more memory implies, once again, that the process is
more gradual (peaks occur at later times), the difference is less
noticeable than in the previous case.

One of the main uses of epidemiological models is to analyze the
effect of possible interventions and to provide useful information for
designing epidemic control strategies
\cite{brauer2012mathematical}. In the context of the present model,
the question arises as to what measures can be employed to combat
radicalization and what are their effects. Given that the behavior of
the system in the long run is governed by the $\Lambda$ and $\Omega$
parameters, and that radicals will have a marginal expression whenever
$\Omega >\Lambda$, , appropriate measures should tend to decrease
$\Lambda$ and/or increase $\Omega$. However, not all the original
parameters of the model are externally controllable. The parameter
$\alpha$ will depend on the level of militancy of the radicals,
$\gamma$ on the level of contentment provided by belonging to such a
group; and $p$ on the efficiency of their convincing strategies as
well as on cultural factors. On the other hand, $\delta$ is a measure
of the probability that a moderate individual will become susceptible,
therefore it will depend on random personal factors and others linked
to social welfare, which can only be modified in the medium-to-long
term\cite{berman2004indicators}. These hypotheses leave as the main
measure the application of policies that encourage positive social
interaction (increasing $\beta$), trying to appease the radical
impulse by depriving it of susceptible individuals.

\begin{figure}[h]
 \centering
 	\includegraphics[width=0.49\columnwidth]{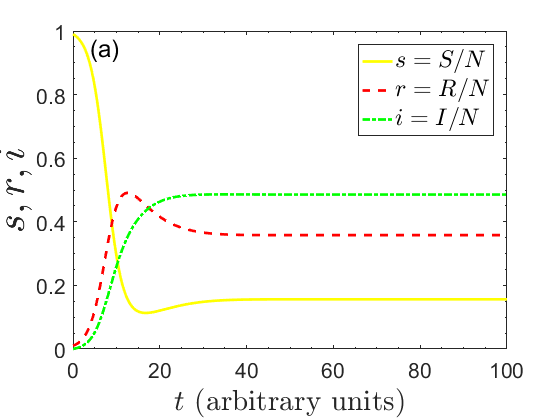}
    \includegraphics[width=0.49\columnwidth]{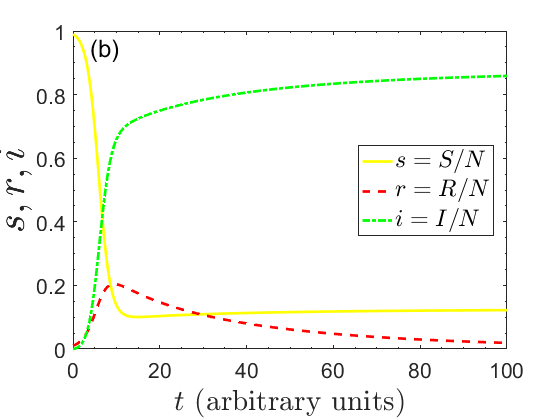}   
 \caption{\textcolor{black}{Evolution of the populations} in the
   presence of a strong radical group (Markovian case),
   \textcolor{black}{without intervention (panel (a)}, $\Omega =0.05$,
   $\Lambda =6.4$); \textcolor{black}{and with a strong intervention}
   (panel (b), $\Omega =8$, $\Lambda =6.4$). The initial distribution
   is $(0.99, 0.01,0)$.}
\label{f:f9}
\end{figure}
As an example, let us consider again the situation corresponding to
case 3, for large values of $\Lambda$. According to the previous
analysis, the population will start at the susceptible state, and if
the system evolves without any intervention, there will be a
significant fraction of radicals in the equilibrium state
\textcolor{black}{(see Fig. \ref{f:f9}a)}. Suppose that, to avoid this,
strong intervention strategies are implemented on the susceptible
population. If $\Omega$ manages to surpass $\Lambda$ (case 5), the
evolution of the populations predicted by the Markovian model will be
similar to that shown in Fig. \ref{f:f9}b. \textcolor{black}{ The
  temporal evolution in this case exhibits a maximum, but with smaller
  amplitude than in the previous case, concurrent with an increase in
  immunized individuals. After overcoming this maximum, the radical
  fraction vanishes in the long term.}

\begin{figure}[h] \centering
	\includegraphics[width=0.49\columnwidth]{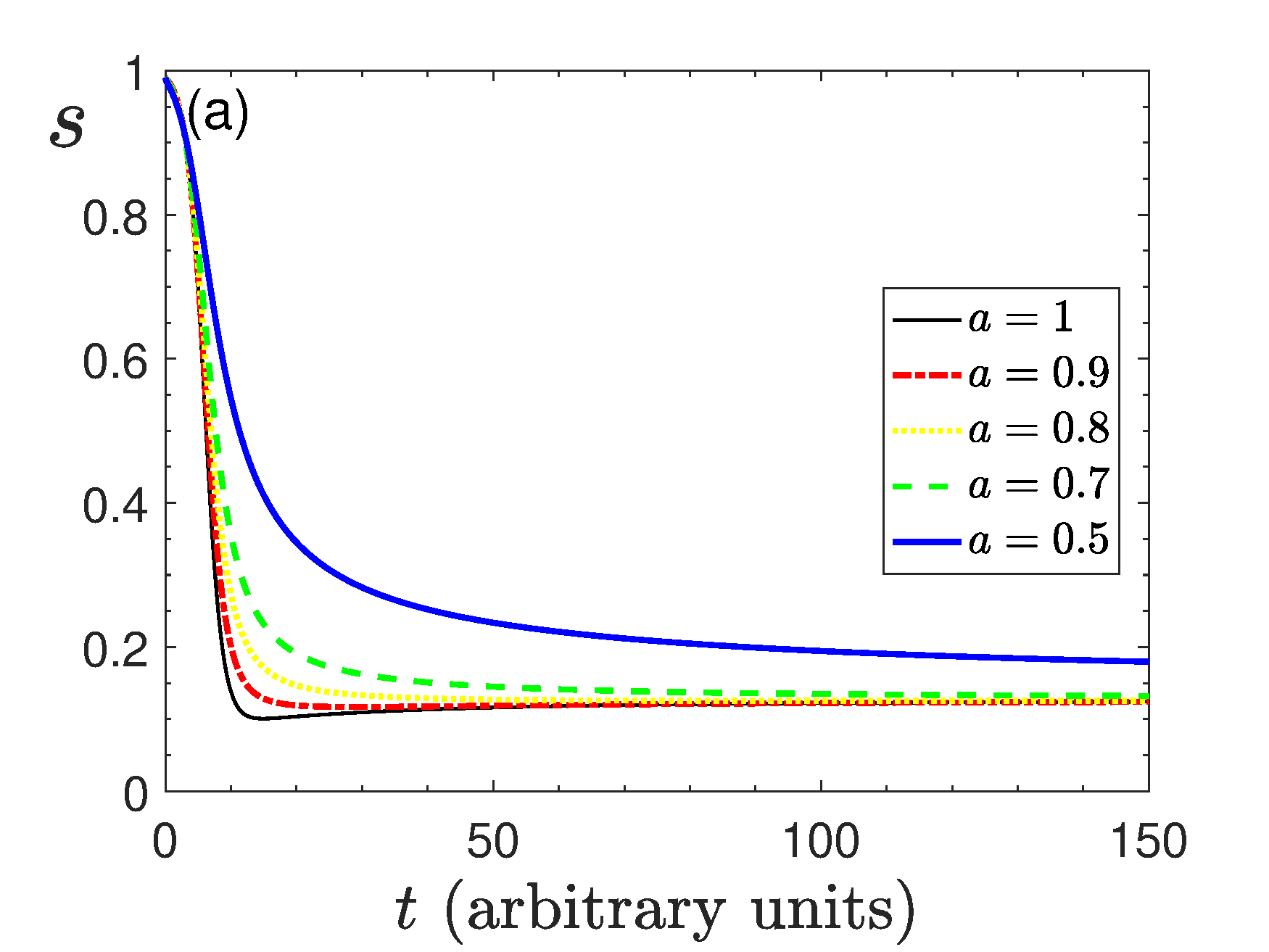}
	\includegraphics[width=0.49\columnwidth]{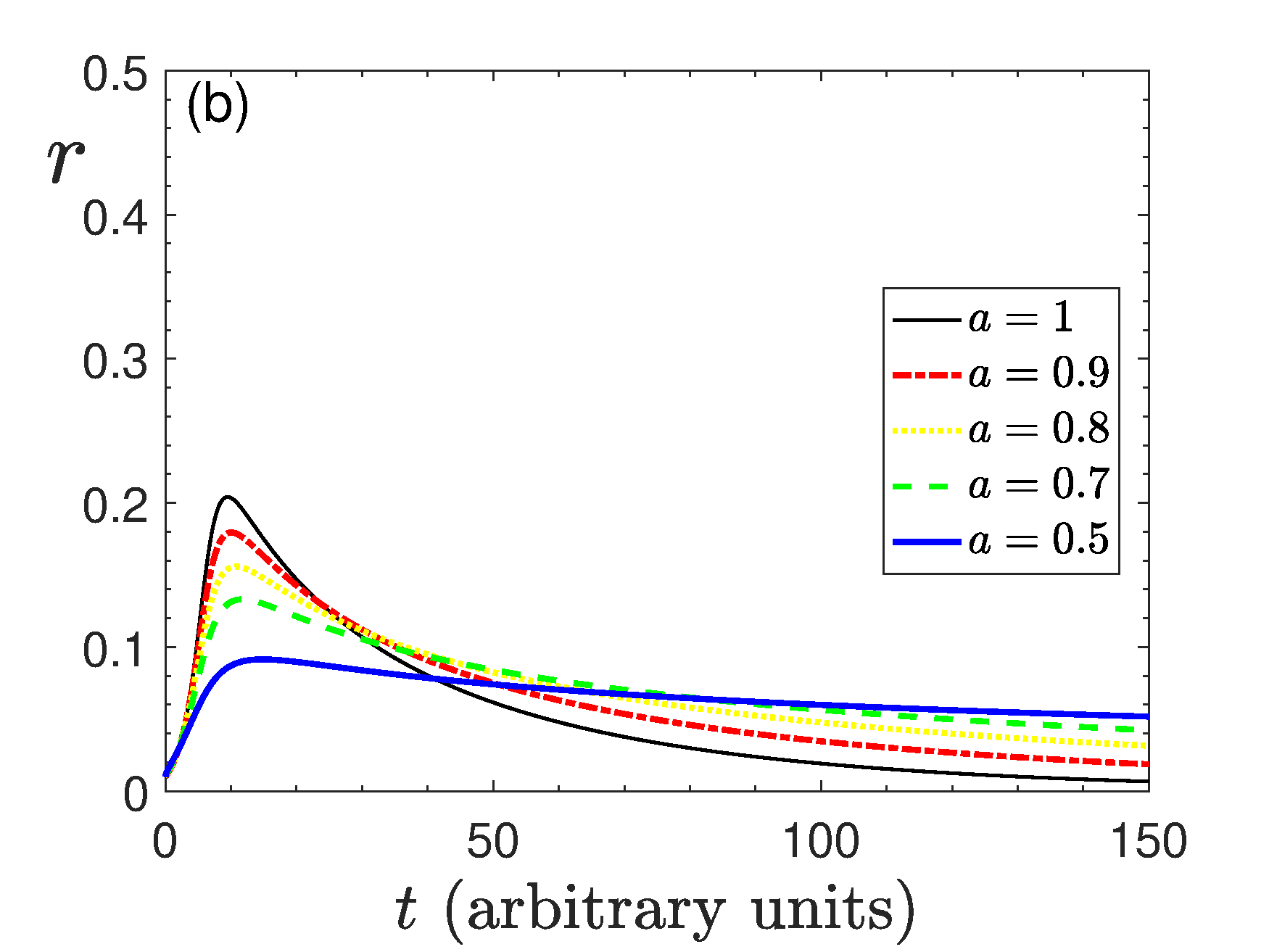}
	\includegraphics[width=0.49\columnwidth]{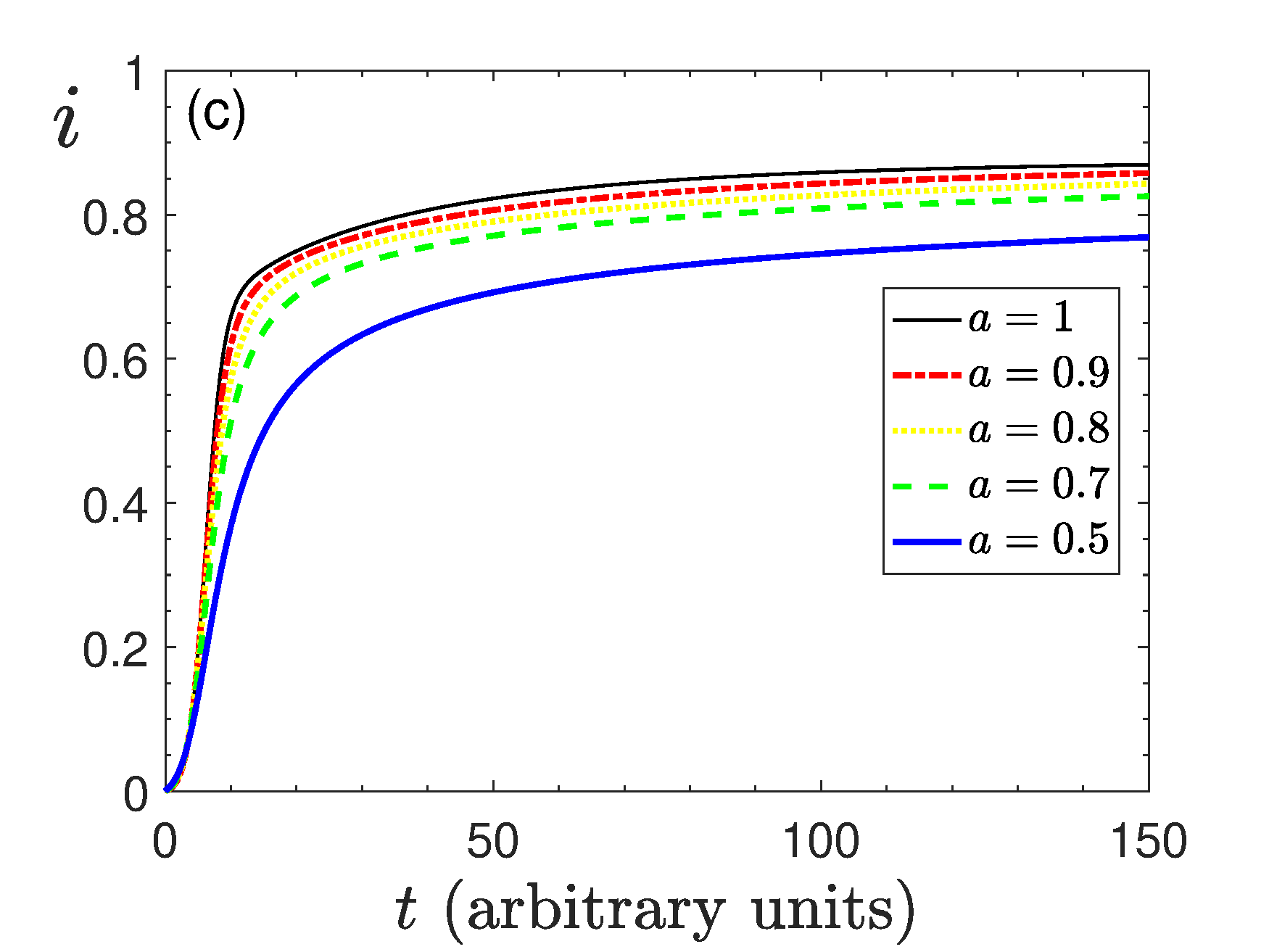}
\caption{Evolution of populations; \textcolor{black}{susceptibles (a),
    radicalized (b) and inmunized (c),} once the interaction with
  susceptibles has been increased for different levels of memory. The
  values of the parameters are: $\alpha =0.8$, $\beta =0.8$, $\gamma
  =0.1$, $\delta =0.1$ and $p=0.8$ ($\Omega =8$, $\Lambda =6.4$). The
  initial population distribution is $(0.99,0.01,0)$.}
\label{f:f10}
\end{figure}

\textcolor{black}{The effects of this policy in societies with different
  levels of memory are shown in Fig.~\ref{f:f10}.  We have verified
  that the increase in positive interaction with the susceptible
  population has the effect described in the above paragraph
  independently of the memory level. However, the characteristics of
  the remaining radicalization peak, as well as of the subsequent
  evolution towards the asymptotic state, strongly depend on the
  parameter $a$.}  \textcolor{black}{Note that, although the time of
  occurrence of the peak is essentially the same in all cases, the
  more memory is included in the system, the smaller the amplitude of
  the peak and the longer the time required to reach an
  equilibrium. For practical purposes, we conclude that the
  application of this policy:} \textcolor{black}{
\begin{itemize}
\item It does not manage to delay the advance of radicalization, in
  the sense that the maximum proportion of radicals is reached at the
  same time as if the policy had not been implemented. Additionally,
  that time is essentially independent of the memory of the system.
\item It is successful in reducing the maximum levels of radicalizaion, for all memory levels.
\item It is successful in the long term, reducing radicalization to
  negligible values.
\item It is more successful in overcoming radicalization in societies
  with short-range memory. The more memory is included in the
  modeling, the longer the coexistence with the radical group.
\end{itemize}}



\section{Conclusion}

In this work we have proposed a simple compartmental model, inspired
by epidemiological models, to describe the dynamics of radical ideas
in a society with memory. The model reasonably predicts under what
conditions the radical group remains in marginal expressions, achieves
a stable presence, or eventually manages to impose itself as the main
expression.

The model depends strongly on two parameters. One of them, $\Lambda$,
represents the number of susceptible individuals that each radical
persuades into joining the group while professing that ideology, thus
representing a measure of the strength of that group. The other
quantity, $\Omega$, plays the same role as $\Lambda$, but for the
\textcolor{black}{immunized} group, and ensures the disappearance of
radicals whenever its value exceeds $\Lambda$. This parameter can be
interpreted from a sociological point of view, noting that the
specific literature agrees that radicalization arises less frequently
in highly cohesive societies. High cohesion implies, on the one hand,
a reasonable level of social welfare related to economic aspects,
public health, low level of inequality and respect for minorities
(linked in our model to the parameter $\delta$); but also to the fact
that all groups that make up society feel duly integrated, the
existence of shared values and a strong interconnection and solidarity
among its members (aspects related to $\beta$). Given that $\Omega$ is
constructed based on these two parameters, we consider that it can
represent a reasonable measure of social cohesion.

The model also highlights the importance of intervention strategies on
vulnerable populations as a tool to control radicalization. The
difficulties faced by deradicalization programs and intelligence
services in combating violent radicalization suggest the need for a
paradigm shift, in which greater importance is given to strategies
aimed at preventing radicalization in its early
stages\cite{grossman2016stocktake,rousseau2017and}. Examples include
interventions in educational settings aimed at youth and
adolescents\cite{tsintsadze2014groupthink, bhui2012public}, or
roundtable discussions to include minority representatives
\cite{keeble201810}. These types of interventions aim at increasing
the interaction between \textcolor{black}{immunized} and susceptible
individuals, so they can be associated with an increase of $\beta$
(and hence of $\Omega$). In the context of this model, we have seen
that these measures, although they do not stop the advance of
radicalization, they flatten the curve and manage to maintain it at
negligible levels in the long term.

The effects of including memory in the modeling of the system plays a
significant role. \textcolor{black}{The level or degree of memory is a
  characteristic of societies over which there is no immediate
  control. However, it is interesting to analyze how the result of the
  intervention policies may be different, depending on the degree of
  memory that a particular society presents}. We observe that at
higher levels of memory, processes occur less abruptly, and it takes
longer to reach a state of equilibrium. In terms of combating
radicalization through interaction with susceptible individuals, the
model predicts that societies with more memory will have lower values
of transient radicalization, but a longer time of coexistence with the
radical group before its disappearance.

\section*{Acknowledgements}
This work was partially supported by Comisi\'on Acad\'emica de
Posgrado (CAP), Agencia Nacional de Investigaci\'on e Innovaci\'on
(ANII) and Programa de Desarrollo de las las Ciencias B\'asicas
(PEDECIBA) (Uruguay).


\bibliography{/home/arturo/Dropbox/bibtex/mybib}

\end{document}